\documentstyle[12pt,aasms4]{article}

\lefthead{Mediavilla et al.}
\righthead{Spectroscopy of the Lens Galaxy of Q0957+561A,B}

\begin{document}

\title{Spectroscopy of the Lens Galaxy of Q0957+561A,B. Implications 
of a possible central massive dark object}

\author{Evencio Mediavilla (1), Miquel Serra-Ricart (1), Alejandro 
Oscoz (1), Luis Goicoechea (2), and Jesus Buitrago (1)} \affil{(1) 
Instituto de Astrof\'\i sica de Canarias, E-38200 La Laguna, 
Tenerife, Spain} \affil{(2) Departamento de F\'\i sica Moderna, 
Universidad de Cantabria, E-39005 Santander, Cantabria, Spain}

\begin{abstract}

We present new long-slit William Herschel Telescope
 spectroscopic observations of 
the lens galaxy G1 associated with the double-imaged QSO 0957+561A,B.
The obtained central stellar velocity dispersion, $\sigma_l = 
310\pm 20\rm\, km\, s^{-1}$, is in reasonable agreement with other 
measurements of this dynamical parameter. Using all updated 
measurements of the stellar velocity dispersion in the internal 
region of the galaxy (at angular separations $<$ 1\farcs5) and a 
simple isotropic model, we discuss the mass of a possible 
central massive dark object. It is found that the data of Falco et 
al. (1997) suggest  the existence of an extremely massive object of
(0.5--2.1)$\times$10$^{10}$ $h^{-1}$ $M_{\odot}$ (80$\%$ confidence 
level), whereas the inclusion of very recent data (Tonry \& Franx 
1998, and this paper) substantially changes the results: the
compact central mass must be $\le$ 6$\times$10$^{9}$ $h^{-1}$ 
$M_{\odot}$ at the 90$\%$ confidence level. We note that, taking 
into account all the available dynamical data, a compact nucleus with 
a mass of 10$^{9}$ $h^{-1}$ $M_{\odot}$ (best fit) cannot be ruled 
out.                                         

\end{abstract}

\keywords{dark matter -- galaxies: elliptical and lenticular, cD -- 
galaxies: structure -- galaxies: kinematics and dynamics -- 
gravitational lensing -- quasars: individual (0957+561)}

\section{Introduction}

Gravitational lenses have turned out to be ideal sites for  
obtaining estimates of several cosmological quantities, such as 
the content of the universe  (see, for example, Im et al. 1997) or 
the Hubble constant. Refsdal (1964a,b) first demonstrated that a 
determination of $H_0$ could be obtained from the time delay between 
two components of a multiply-imaged quasar and an estimate of the 
distribution of mass in the deflector. This method is based on 
purely geometrical grounds (within the framework of the general 
theory of relativity), and presents decisive advantages with 
respect to  conventional astronomical techniques (see, for example, 
Kundi\'c et al. 1997). In particular, it avoids  uncertainties 
in the construction of the cosmological distance ladder, and 
 allows the expansion rate from high-redshift---i.e., 
cosmologically distant---objects to be obtained.

Q0957+561A,B was the first gravitational lens system to be 
discovered (Walsh, Carswell, \& Weymann 1979), and has been 
monitored since then (see, for example, Schild \& Thompson 1995; 
Oscoz et al. 1996). However, despite  the several observational 
campaigns carried out, the value  for the time delay was not
well determined until very recently (Kundi\'c et al. 1997; Oscoz 
et al. 1997). Nowadays, a time  delay close to 420 days is widely 
accepted. Furthermore, it is  known that the main gravitational 
effect in Q0957+561A,B is 
produced by two intervening objects: a giant elliptical galaxy 
(Brightest Cluster Galaxy, G1), discovered by Stockton (1980), 
and a cluster of galaxies  surrounding G1. Several mass models, 
in which the form of the lens  (galaxy + cluster) profile were 
discussed, have been proposed since 1979 (Young et al. 1980; 
Borgeest \& Refsdal 1984; Greenfield et al. 1985; Kochanek 1991; 
Falco et al. 1991; Grogin \& Narayan 1996a,b; Barkana et al. 1998). 
Unfortunately, the 
models are incomplete (the fits are unable to discriminate
 the contributions from mass associated with the galaxy 
and mass associated with the cluster), and are therefore unable
to compute the Hubble constant without additional information. 
To break this degeneracy, a measurement of either the velocity 
dispersion of G1 (the 1D velocity dispersion associated with the 
dark objects in the halo of G1) or the surface mass density of  
the cluster (Fischer et al. 1997) is needed. 

The first attempt to measure $\sigma_{l}$ (the 1D velocity 
dispersion associated with the luminous stars in G1) was made by 
Rhee (1991) who, analyzing the Mg~{\sc i}{\it b} triplet absorption 
lines from the spectra obtained with the 4.2-m William Herschel 
Telescope (WHT), inferred a value for the stellar velocity 
dispersion of G1 of $\sigma_{l} = 303 \pm 50$ km s$^{-1}$. Some 
years later, Falco et al. (1997, hereafter FSMD) have also 
measured the Mg~{\sc i}{\it b} triplet with the Keck Telescope, 
reporting an apparent decrease of $\sigma_{l}$ with angular distance 
from the center of G1 (from 316 km s$^{-1}$ at $\varphi \approx$ 
0\farcs1 to 262 km s$^{-1}$ at $\varphi \approx$ 0\farcs5) and 
obtaining an averaged value of $279\pm 12\rm\, km\, s^{-1}$. Any 
giant elliptical could have a central black hole of 10$^{7}-$10$^{9}$ 
$M_{\odot}$ (see Kormendy et al. 1998), and so the gradient has 
been preliminarly interpreted by the authors as due to a central 
massive dark object of mass $\approx$ 4$\times$10$^{9}$ $h^{-1}$ 
$M_{\odot}$ ($H_{0}$ = 100 $h$ km s$^{-1}$ Mpc$^{-1}$), i.e., a mass 
similar to the largest measured central mass in elliptical 
galaxies. By means of a more careful analysis (see \S 
4.2), it can be shown that the decrease of $\sigma_{l}$ with 
distance to the galaxy center reported by FSMD should imply a 
central object with a very large mass of $\approx$ 10$^{10}$ 
$h^{-1}$ $M_{\odot}$ (assuming, as do FSMD, that the gradient is 
only due to a compact nucleus). Very recently, Tonry \& Franx 
(1998, hereafter TF) have measured (also with the Keck Telescope) 
the central stellar velocity dispersion inside an aperture of 
1\arcsec $\times$ 2\farcs3 as well as the $\sigma_{l}$ from 
different rows along the spectrograph slit. The slit was rotated by 
90 deg with respect to the previous orientation by FSMD, and with 
this new configuration, no gradient is seen in the stellar velocity 
dispersion trend. Therefore, TF do not agree with the conclusion 
of FSMD concerning the existence of a central dark object with very 
large mass. In fact, their dynamical data are in apparent agreement 
with the absence of a compact nucleus.      

We present here a new measurement of $\sigma_l$ (the 
spectrograph slit was rotated by about 10 deg and 20 deg
with respect to the exposures 1$-$3 and 4 by FSMD, respectively)
of similar accuracy to those of FSMD and TF. In \S 2 we present the 
observations and data reduction. Section 3 is devoted to the 
derivation of $\sigma_{l}$ ($\varphi$  $\le$ 1\farcs5) from the 
spectra, and in \S 4 we use all the updated information (the data of 
FSMD and TF, and ours) to discuss the mass of a possible central
massive dark object. 

\section{Observations and Data Reduction}

The spectra of the lens galaxy in Q0957+561 were collected at 
the WHT (La Palma, Canary Islands, Spain) on 1997 February 4. 
We used the red arm of the Intermediate-dispersion Spectrograph 
and Imaging System (ISIS) with a Tek CCD camera. The 158 line 
mm$^{-1}$ grating, R158R, gave a dispersion of 1.21 \AA\  
pixel$^{-1}$, covering a spectral range from 5000 to 8000 \AA. 
The seeing was 1$\arcsec$ FWHM and the spectral 
resolution, derived from comparison lamps, was $\sim 2.2$ \AA. 
We centered a 1\arcsec $\times$ 240$\arcsec$  
slit on G1 (but  an aperture of only 1\arcsec $\times$ 
3$\arcsec$ is used), and oriented it along P.A. 
$\sim 190^\circ$ to  include image B also in the aperture (see 
Fig. 1). We took 11 exposures of 1800 s each. We also obtained 
three exposures of 10 s  of the calibration star AGK 2+14783.

Data reduction was carried out using the standard software 
package {\sc iraf}\footnote{{\sc iraf} is distributed by the 
National Optical Astronomy Observatories, which is operated by 
the Association of Universities for Research in Astronomy, Inc. 
(AURA) under cooperative agreement with the National Science 
Foundation.}. The spectrum images were bias-subtracted and trimmed, 
flat-field corrected and cleaned of cosmic rays. Wavelength 
calibration and distortion correction were performed with 
polynomial fits to the Cu--Ne arc lamp spectra. Finally, we co-added 
all the 1800-s spectra. In Fig. 2 we show the resulting spectrum. 
In the analysis of the next section we will also make use of 
some data from the calibration star HD~27697 kindly provided by 
Peletier et al. (1997), obtained in the same spectrograph but with
higher spectral resolution.

\section{Results}

Prior to computing the FWHM of the Mg~{\sc i}{\it b} lines, the 
spectrum of the lens galaxy had to be continuum normalized and 
subtracted. First  the continuum was drawn using as reference the 
calibration-star spectra and the results of FSMD. We fitted the 
continuum points to splines with CDRAW, a command of DIPSO (Howarth 
\& Murray 1988). Next, the spectrum was divided by the 
fitted continuum. Finally, the continuum level was set to zero by 
subtracting 1. The resulting spectrum (blueshifted dividing 
$\lambda$ by $1+z=1.3569$) in the wavelength range of the 
Mg~{\sc i}{\it b} triplet is shown in Figure 3.

To obtain a first estimate of the FWHM we  attempted a three-Gaussian 
fit (constraining the  central lambdas of the Gaussian functions by 
using the atomic wavelengths) to the Mg~{\sc i}{\it b} lines 
corresponding to G1. We have supposed the same width for the three 
lines and fitted another two Gaussian functions to the spectral 
absorption features close to the triplet in the red and blue wings, 
respectively. To restrict the fitting further, we have  constrained 
the relative intensities of the lines using as reference values the 
quotients 0.85:1:0.81, found fitting the Mg~{\sc i}{\it b} triplet of
 the calibration star AGK 2+14783. 
After correcting for the instrumental width we obtain a value of 
$\sigma_l=323\pm 19\rm\, km\, s^{-1}$. The uncertainty 
represents the rms error of the multi-Gaussian fit to the data but 
does not include systematic sources of error coming from the use of 
the reference values for the relative intensity of the lines, or 
from the continuum correction. The last is a very important source 
of error since the galaxy absorption lines are strongly diluted by 
the QSO emission.

To obtain more confident values for $\sigma_l$, we  used 
the standard cross-correlation technique, in the following steps: 
(i) continuum correction (as described above) of the galaxy and 
template spectra, (ii) transformation of the wavelength axis 
taking logarithms, and (iii) cross-correlation of the spectra 
with an appropriate template using the command XCORR of DIPSO, 
which interpolates the data and masks the edges of each spectrum 
with a cosine bell. As templates we  used alternatively: AGK 
2+14783 observed on the same night as the galaxy under the same 
instrumental conditions, HD~27697 observed with higher resolution, 
and this last spectrum convolved with a Gaussian of 6.58\AA\ 
(FWHM) to match the resolution of the galaxy observations. To 
derive $\sigma_l$ from the galaxy spectrum/stellar template
cross-correlatation function, we have used the standard method 
 also followed by FSMD.

This consists in fitting a Gaussian to the central points of the
cross-correlation function peak. The FWHM of the Gaussian fit 
depends on the template and galaxy broadenings. To 
find the FWHM($\sigma_l$) relationship the method is calibrated
by cross-correlating the template with itself smoothed with Gaussian
functions
of different widths. In Fig. 4 we present the FWHM($\sigma_l$) 
calibrations corresponding to the three template
used by us. Using this method we have obtained the following 
$\sigma_l$ values corresponding to each template: $310\pm 
17\rm\, km\, s^{-1}$ (template: AGK 2+14783); $310\pm 27\rm\, 
km\, s^{-1}$ (HD27697); and $331\pm 35\rm\, km\, s^{-1}$ (HD27697 
convolved with a Gaussian of FWHM = 6.58\AA\ ). The uncertainties 
in the previous computations correspond to the rms errors in the
cross-correlation fitting width. The best estimate corresponds 
to the star observed on the same night (AGK 2+14783), which seems 
reasonable. Since there are no  significant differences with the 
results obtained with the other templates, we adopt it as the 
final value. In Fig. 5 we present a smoothed ($\sigma=310\rm\, 
km\, s^{-1}$) spectrum of the calibration star AGK 2+14783 
superimposed on the 0957+561 spectrum. To estimate the global 
uncertainty, we consider two different sources of error: (i) 
the rms error in the cross-correlation fitting width ($17 \rm\, 
km\, s^{-1}$), and (ii) the selection of the template star. The 
dispersion among the three results derived using the three 
different templates ($\sim 10\rm\, km\, s^{-1}$) is a 
conservative estimate of (ii). Consequently, we derive a total 
uncertainty of $\sim 20\rm\, km\, s^{-1}$.

The value for the stellar velocity dispersion of the lens galaxy 
derived by us (310 $\pm$ 20 km s$^{-1}$) is in agreement with the 
value of 303 $\pm$ 50 km s$^{-1}$ obtained by Rhee (1991), but 
there is a difference of 31 $\pm$ 23 km s$^{-1}$ between the new 
result and the averaged value of 279 $\pm$ 12 km s$^{-1}$ 
presented by FSMD. To exclude the possibility that this
difference arises from the data analysis, we  repeated the 
computations following the procedure used by FSMD and obtained 
essentially the same results. There is a relatively good agreement 
with respect to the central stellar velocity dispersion by TF (288 
$\pm$ 9 km s$^{-1}$). The 
difference of 22 $\pm$ 22 km s$^{-1}$ is consistent with zero.

\section{Discussion}

\subsection{A Simple Isotropic Model}

To  relate physically the derived values for $\sigma_l$ (FSMD, TF 
and this paper) with the compact central mass of G1 we focused on 
a simple scenario for the galaxy: a singular isothermal sphere, 
SIS, with velocity dispersion $\sigma_{h}$, plus a central massive 
dark object, MDO, with mass $M_{c}$. The halo of luminous stars ($\rho_{l}$) 
and the halo of dark objects ($\rho_{m}$) can be considered as 
spherically symmetric ideal gas spheres (the velocity distribution 
is isotropic). In that case, the gas pressure is
\begin{equation}
p_{i} = \rho_{i}\sigma_{i}^2 ,
\end{equation}
and the equation of hydrostatic equilibrium,
\begin{equation}
dp_{i}/dr = -[GM(r)/r^{2}]\rho_{i} , 
\end{equation}
with $i=l,m$ and $M(r) = 4\pi\int_{0}^{r} \rho_{m}r^{2}dr$. For 
a SIS model of dark matter, $\rho_{m} = \sigma_{h}^{2}/2Gr^{2}\pi$ 
and $\sigma_{m} = \sigma_{h}$, a velocity dispersion independent 
of the radius. However, the inclusion of a compact nucleus 
(SIS+MDO), $M(r) = M_{c} + 2\sigma_{h}^{2}r/G$, leads to
$\sigma_{m}^{2} = \sigma_{h}^{2} + GM_{c}/3r$, and now $\sigma_{m}$ 
is not constant at every point of the galaxy. To obtain 
$\sigma_{l}$ it is necessary to know the behavior of $\rho_{l}(r)$.
This information can be inferred by considering that the observed
surface brightness profile, $I(s)$, traces the surface density of
luminous matter, 
\begin{equation}
I(s) \propto \Sigma_{l}(s) = 2\int_{s}^{\infty}
\frac{\rho_{l}(r)rdr}{(r^{2}-s^{2})^{1/2}}     . 
\end{equation}

{\it Hubble Space Telescope (HST)} data indicate that $I(\varphi) 
\propto \varphi^{-n}$, $n \approx$ 1 (0\farcs1 $\leq \varphi 
\leq$ 5\arcsec), where the $\varphi-s$ relationship for G1 
($\Omega_{0}$ = 1) is given by $\varphi(^{\prime\prime}) \approx 
(1/3)$h$^{-1}$$s$(kpc), although the whole surface brightness profile 
is not well fitted by a single power law (Bernstein et al. 
1997). At the {\it HST} resolution limit, the light 
distribution could be singular, and 
so, a singular isothermal halo ($\rho_{l} = Ar^{-2}$) would be in rough
agreement with the {\it HST} profile. Our master equation will be 
\begin{equation}
\sigma_{l}^{2} = \sigma_{h}^{2} + GM_{c}/3r .
\end{equation}
We admit that the previous relations (1--4) are valid at 
separations ($s$) and radii ($r$) much greater than the 
Schwarzschild radius of the mass $M_{c}$. For example, taking 
$M_{c}$ = 4$\times$10$^{9}$ $h^{-1}$ $M_{\odot}$, the Schwarzschild 
radius is of about 4$\times$10$^{-4}$ $h^{-1}$ pc. Therefore, $R$ = 1
$h^{-1}$ pc ($\approx$ 1/3 mas in angular separation) can be a 
reasonable threshold of validity (we avoid a complex description
of the region at $r < R$, which should include relativistic 
effects, the probable existence of an accretion disc, the possible 
distortion of the luminous isothermal halo as due to ``cannibalism" 
by the system central massive dark object + accretion disc, etc.).

In practice, one does not measure the stellar velocity dispersion, 
$\sigma_{l}(r)$, but rather either the line-of-sight stellar 
velocity dispersion  at separation $s$ (using a small aperture) 
or the velocity dispersion associated with the luminous stars
inside a finite aperture, $S$. In order to compare with 
observations, the useful expressions are
\begin{equation}
\sigma_{l}^{2}(S) = \frac{\int_{S} 
\sigma_{l}^{2}(s)\Sigma_{l}(s)dS}{\int_{S} \Sigma_{l}(s)dS}   , 
\end{equation}
and
\begin{equation}
\sigma_{l}^{2}(s) = \sigma_{h}^{2} + 2GM_{c}/3s\pi ,
\end{equation}
where $\Sigma_{l}(s) \propto s^{-1}$. If the ``forbidden" central
circle ($\pi R^{2}$, where $R$ is the threshold of validity; see 
comment after Eq. (4)) is included in $S \gg \pi R^{2}$, we assume 
that its contribution to $\sigma_{l}(S)$ is negligible. Equations 
(5--6) are  applicable only when there is no seeing, and must be 
properly modified accounting for seeing effects in any realistic
case; e.g.,
\begin{equation} 
\sigma_{l}^{2}(s,S_{\ast}) = \frac{\int_{S_{\ast}} 
\sigma_{l}^{2}(t)\Sigma_{l}(t)P({\bf t},{\bf s})d^{2}{\bf t}}  
{\int_{S_{\ast}} \Sigma_{l}(t)P({\bf t},{\bf s})d^{2}{\bf t}} =  
\sigma_{h}^{2} + (2GM_{c}/3\pi)(\lambda_{2}/\lambda_{1}) , 
\end{equation} 
where $P({\bf t},{\bf s}) \propto \exp[-({\bf t}-{\bf  
s})^{2}/2s_{\ast}^{2}]$ represents the point spread function 
corresponding to a seeing of Gaussian dispersion $s_{\ast}$ and
$\lambda_{q} = \int_{S_{\ast}} t^{-q}P({\bf t},{\bf s})d^{2}{\bf 
t}$ ($q$=1,2). Equation (7) is the realistic version of Eq. (6).
It is now relatively easy to estimate $\lambda_{q}$ ($q$=1,2) at 
$s < 2s_{\ast}$, and so, to infer
\begin{equation}
\sigma_{l}^{2}(s,S_{\ast}) = \sigma_{h}^{2} + 
\sqrt{8/9\pi^{3}}(GM_{c}/s_{\ast})
(1+\frac{s^{2}}{4s_{\ast}^{2}})^{-1} \ln{(2s_{\ast}/R)} .
\end{equation}
From this equation, at $s < s_{\ast}$, one concludes that the 
``heating" [$\Delta T \propto \sigma_{c}^{2} = 
\sigma_{l}^{2}(s,S_{\ast}) - \sigma_{h}^{2}$] caused by the 
compact nucleus is not traced by a  $1/s$ law [as  derived 
from Eq. (6)]. On the contrary, at this region will appear a 
quasi-isothermal behaviour due to the Gaussian smoothing of the 
seeing. As the angular separations $\varphi$ are 
proportional to the separations $s$, we can rewrite $\sigma_{c}$ 
(in km s$^{-1}$) and 
\begin{equation}
\sigma_{l}^{2}(\varphi,\sigma_{\ast}) = \sigma_{h}^{2} + 
2.4\times10^{-7}(hM_{c}/\sigma_{\ast})(1+\frac{\varphi^{2}}
{4\sigma_{\ast}^{2}})^{-1}\ln{(6\times10^{3}\sigma_{\ast})} ,
\end{equation}
where $hM_{c}$ is in $M_{\odot}$, and $\varphi$ and $\sigma_{\ast}$
[$\sigma_{\ast} = \varphi(s_{\ast})$] are in  arcsec. In this paper,
we concentrate on an isotropic model incorporating a compact 
central mass, $M_{c}$, which is responsible for a gradient in the 
stellar velocity-dispersion trend.

\subsection{Application to Real Data}

Using the data derived by FSMD, equation (9) allows us to make a 
first isotropic model measurement of $hM_{c}$ (isotropic model 
measurements of the mass at the center of the galaxies are usual, 
e.g., Kormendy et al. 1998). We have estimated the angular 
separation $\varphi(i)$ ($i$=1,5) corresponding to the centers of 
the five rows (small rectangles of 1\arcsec $\times$ 
0\farcs2) used by FSMD, ranging from $\varphi(3)$ = 
0\farcs21 to $\varphi(5)$ = 0\farcs53. As 
the spectra were taken with a $0\farcs8$ (FWHM) mean seeing (see 
Table 1 of FSMD), at $\varphi$ $^{<}_{\sim}$ 0\farcs5, the law (9) 
can be applied ($\sigma_{\ast}$ = 0\farcs34). Avoiding the finite 
size of the rectangles (due to seeing, it is expected that the 
inhomogeneities in the rows will be small), one can establish a  
relationship between the observations $\sigma_{l}(i)$ and 
[$\varphi(i),\sigma_{\ast}$] via Eq. (9). Our model fits the five 
constraints with two free parameters, i.e., we have 3 degrees of 
freedom ($dof$ = 3). The best-fit model gives $\chi^{2}/dof 
\approx$ 3.5, and due to this poor $\chi^{2}/dof$, we have drawn 
the $\Delta(\chi^{2}/dof)$ = 1--10 contours rather than those for 
$\Delta\chi^{2}$ = 1,2,...  (see Fig. 6). Our best-fit 
velocity dispersion is  $\sigma_{h}$ = 39 km s$^{-1}$, and 
models with $\sigma_{h} \geq$ 250 km s$^{-1}$ correspond to  
$\Delta(\chi^{2}/dof) >$ 2. This is a first surprising result.
However, the central MDO has a mass in the interval: 
(0.5-2.1)$\times$10$^{10}$ $h^{-1}$  
$M_{\odot}$ [$\Delta(\chi^{2}/dof)$ = 2 limits]. We consider the
1$\sigma$, 2$\sigma$,... bounds such that they correspond to
$\Delta\chi^{2}$ = $\chi^{2}/dof$, $\Delta\chi^{2}$ = 
4($\chi^{2}/dof$),... rather than $\Delta\chi^{2}$ = 1,4,... (see, 
for example, Grogin \& Narayan 1996a; Barkana et al. 1998), and 
consequently, here, $\Delta(\chi^{2}/dof)$ = 2 limits are 
1.3$\sigma$ limits (80$\%$ confidence limits). So, the  FSMD data 
agree with the existence of a very large compact central mass in G1. 

When the aperture $S$ is large, we must use a modified version of 
Eq. (5) instead of Eq. (7). In the presence of seeing, the velocity 
dispersion associated with the stars inside a finite aperture, $S$, 
is given by 
\begin{equation}
\sigma_{l}^{2}(S,S_{\ast}) = 
\frac{\int_{S}dS 
\int_{S_{\ast}} \sigma_{l}^{2}(t)\Sigma_{l}(t)P({\bf t},{\bf
s})d^{2}{\bf t}} 
{\int_{S}dS \int_{S_{\ast}} 
\Sigma_{l}(t)P({\bf t},{\bf s})d^{2}{\bf t}} = \sigma_{h}^{2} + 
(2GM_{c}/3\pi) 
[\frac{\int_{S}\lambda_{2}dS}{\int_{S}\lambda_{1}dS}] .
\end{equation}
As was shown in \S 3, we have presented a
new measurement of the stellar velocity dispersion. The mean 
seeing was $1''$ (FWHM), and the aperture was a rectangle 
centered on G1 with semisides $\rm FWHM/2=0\farcs5$ and $\rm 
3FWHM/2=1\farcs5$. Under these observational conditions, it is 
possible to show that Eq. (10) can be rewritten as Eq. (9), by
using an effective angular separation of 
2$\sigma_{\ast}/\sqrt{3}$. So, if our measurement is labeled with 
the number ``6", $\sigma_{l}$(6) will be compared to 
$\sigma_{l}$[$\varphi$(6) = 2$\sigma_{\ast}/\sqrt{3}$, 
$\sigma_{\ast}$] = $\sigma_{l}$(0\farcs48, 
0\farcs42). Re-evaluating $\chi^{2}/dof$ with the six 
constraints ($dof$ = 4), we obtain the contour plots appearing in 
Figure 7. The best-fit $\chi^{2}/dof$, velocity dispersion and central 
mass are now $\approx$ 4, 161 km s$^{-1}$ and 1.37$\times$10$^{10}$ 
$h^{-1}$ $M_{\odot}$, respectively. Now, adding the constraint 
inferred from our spectroscopic study, 
 a shift of the solutions towards lower masses and larger 
values of $\sigma_{h}$ appears.

Finally, we  measure $hM_{c}$  using all the  
dynamical information at $\varphi$  $<$ 1\farcs5: 5 constraints 
from FSMD + 1 constraint from this paper + 4 constraints from TF. 
TF have obtained the central stellar velocity dispersion 
associated with a spectrum which was averaged over 11 rows 
(an aperture of 1\arcsec $\times$ 2\farcs3) along the slit. Taking 
into account a point spread function characterized by an FWHM 
$\approx$ 0\farcs8 (see TF), the aperture can be approximated as 
a rectangle with semisides FWHM/2 and 3FWHM/2. Then, the averaged 
measurement $\sigma_{l}$(7) must be compared to 
$\sigma_{l}$[$\varphi$(7) = 2$\sigma_{\ast}/\sqrt{3}$, 
$\sigma_{\ast}$] = $\sigma_{l}$(0\farcs39, 0\farcs34). However,
 from the results by TF, stellar 
velocity dispersions of 289.5 $\pm$ 7 km s$^{-1}$ at $\varphi$(8) 
= 0\farcs0, 297 $\pm$ 10 km s$^{-1}$ at $\varphi$(9) = 0\farcs6,
and 290.5 $\pm$ 7 km s$^{-1}$ at $\varphi$(10) = 0\farcs7 can be
derived. We 
note that Eq. (9) works relatively well even at $\varphi \approx 
2\sigma_{\ast}$, and so this analytical law is also applied to 
the last two  dynamical data. The new contour plots are shown in 
Fig. 8, and the new best-fit $\chi^{2}/dof$ ($dof$ = 8), velocity 
dispersion and central mass are $\approx$ 3, 283 km s$^{-1}$ and 
10$^{9}$ $h^{-1}$ $M_{\odot}$, respectively. With the whole 
data set, $M_{c} \le$ 6$\times$10$^{9}$ $h^{-1}$ $M_{\odot}$ at 
the 1.63$\sigma$ (90$\%$) confidence level, i.e., a result in 
clear disagreement with the initial one (Fig. 6) based on the data of 
FSMD data. Therefore, we conclude that the existence of a ``normal''
MDO with a mass of about 10$^{9}$ $h^{-1}$ $M_{\odot}$ rather than a 
very large compact central mass is favored by the combined (FSMD + 
TF + this paper) data set.

The predicted values of the stellar velocity dispersions for the 
final best-fit (solid line) and the best-fit to FSMD data
(dashed line) are compared to the observational values 
(1$\sigma$ limits) in Fig. 9. Filled circles, open circles and an 
open square represent FSMD data, TF data and the new measurement, 
respectively. The trend observed by TF (open circles) and the 
dynamical value derived in Sect. 3 (the open square) clearly 
disagree with the solution corresponding to the FSMD data (dashed 
line). From another point of view, three data by FSMD are in 
disagreement with the final fit (solid line) and the data by TF 
and us. Thus, there is evidence in favour of either some kind 
of error in several measurements or an underestimate of several 
error bars. New studies must throw light on this problem and lead 
to a better reduced chi-squared.

\subsection{Some Refinements}

In the two previous subsections, we assumed an isothermal profile
for the luminous tracers. However, Fig. 10 gives the observational
surface brightness (filled triangles; Bernstein et al. 1997) and 
the results of the profile fitting (solid and dashed lines). If 
$I(\varphi) \propto \varphi^{-n}$, the data at $\varphi$  $\le$ 
0\farcs2 show a slope $n$ = 0.65$\pm$0.04 ($\chi^{2}/dof \approx$
1), while the profile is well fit by a power law with $n$ = 1.26
$\pm$0.01 ($\chi^{2}/dof \approx$ 1) in the interval [0\farcs2,   
5\arcsec]. The whole {\it HST} trend can be also fitted by a single
power law: $n$ = 1.12$\pm$0.01. This global fit (solid line) is not
so well as the partial ones (dashed lines) and suggests the existence
of a slight global deviation from the isothermal behaviour.

The switch from $n$ = 1 (isothermal) to $n$ = 1.12 may have some 
effect on the aperture-seeing integrals that appear in the definition
of $\sigma_{l}^{2}(S,S_{\ast})$ [see Eq. (10)]. For 1 $\le n \le$ 2, 
one has $\Sigma_{l}(t) \propto t^{-n}$ and an interpolated analytical
dynamical law
\begin{equation}
\sigma_{l}^{2}(t) = \frac{2}{n+1} \sigma_{h}^{2} + 
\frac{n+1}{(n+2)^2} (GM_{c}/t) .
\end{equation}
From Eq. (11) and $\Sigma_{l}(t) \propto t^{-n}$ ($n$ = 1.12), and 
taking the aperture and seeing conditions associated with our 
measurement, we can recalculate the ratio between the two 
aperture-seeing integrals. It is clear that the factors $1/(n+1)$ and
$(n+1)/(n+2)^2$ in $\sigma_{l}^{2}(t)$ are almost insensitive to the 
change from $n$ = 1 to $n$ = 1.12 (they vary in 2-6\%). On the contrary,
an accurate numerical estimate indicates a significant change of about
50\% in the term related to the central mass $M_{c}$. This last result
is due to the use of a brightness profile ($\Sigma_{l} \propto 
t^{-1.12}$) more realistic than the isothermal profile.

The high sensitivity of $\sigma_{l}(S,S_{\ast})$ (in the term 
proportional to $M_{c}$) to the light distribution $\Sigma_{l}$, forces
us to recompute the central mass. We now use the global value $n$ = 1.12
in $\sigma_{l}$, the double behaviour $n$ = 0.65 at $\varphi$  $\le$ 
0\farcs2 and $n$ = 1.26 at $\varphi$  $\ge$ 0\farcs2 in $\Sigma_{l}$, and
accurate numerical integrals for the ten observational apertures. The
new measurements of the central mass are in excellent agreement with the
"rough measurements" quoted in subsection 4.2. Thus, the FSMD data lead to
$\sigma_{h} \le$ 224 km s$^{-1}$ and $M_{c} = (3.5^{+ 0.3}_{- 2.0})
\times 10^{10} h^{-1} M_{\odot}$ (1$\sigma$ limits), and considering all
data, we infer (1$\sigma$ limits) $\sigma_{h}$ = 294$^{+ 9}_{- 20}$ km 
s$^{-1}$ and $M_{c} \le$ 6.2$\times$10$^{9}$ $h^{-1} M_{\odot}$. Moreover,
the new values of $\chi^{2}/dof$ (best-fits) are very similar to the
old ones.
 
\section{Conclusions}

We have presented new spectroscopy of the lens galaxy (G1) of QSO 
0957+561A,B and have determined the stellar velocity dispersion 
integrated in a rectangular square of $1''\times3''$ centered on 
the galaxy. The obtained value, $\sigma_l=310\pm 20\rm\,km\,s^{-1}$, 
is somewhat greater than the determination by FSMD 
($279\pm12\rm\,km\,s^{-1}$), but in agreement with the measurement  
of $288\pm9\rm\,km\,s^{-1}$ that TF have published during the
preparation of this paper.

Motivated by the controversy about the amount of compact dark 
matter at the center of G1 (a mass of 4$\times$10$^{9}$ $h^{-1}$ 
$M_{\odot}$ was claimed by FSMD for the central object, whereas TF 
discounted the presence of a compact nucleus), we have studied 
the mass of the compact nucleus in G1 by means of a simple 
isotropic model (SIS, singular isothermal sphere, plus an MDO) and 
the available measurements at angular separations $<$ 1\farcs5 
(FSMD, TF, and this paper). Using exclusively the data of FSMD, 
one obtains a surprising best-fit mass measurement of about 
2$\times$10$^{10}$ $h^{-1}$ $M_{\odot}$ for the object in the 
center of the galaxy. So we find that the previous rough estimate 
by FSMD could differ by almost an order of magnitude with respect 
to the value derived from the SIS+MDO model. At the 80\% confidence 
level the compact central mass should be in the range 
(0.5-2.1)$\times$10$^{10}$ $h^{-1}$ $M_{\odot}$. This result would 
be a challenge to current knowledge concerning compact nuclei in galaxies 
(see, for example, fig. 17 of Kormendy et al. 1998). When we 
consider our measurement in the fit, the solution shifts towards 
lower masses with a best fit of 1.4$\times$10$^{10}$ $h^{-1}$ 
$M_{\odot}$. More significantly, adding this new constraint the 
80\% confidence level includes ``normal" MDOs (10$^{7}$--10$^{9}$ 
$M_{\odot}$) and even the absence of a compact central mass. 
To complete the analysis, we also incorporated the dynamical data 
by TF. The final fit (using ten different measurements) suggests 
the existence of an MDO with a mass of about (best-fit) 10$^{9}$ 
$h^{-1}$ $M_{\odot}$. At the 90\% confidence level the compact 
central mass inferred from the ten measurements must be $\le$ 
6$\times$10$^{9}$ $h^{-1}$ $M_{\odot}$. Finally, we remark that 
the measurements of the central mass are based on the assumption 
of an isothermal surface density of luminous matter. However, a
refined model including a more realistic $\Sigma_{l}$ (faithfully
traced by the observational brightness profile) led to very similar
results (see subsection 4.3).                                   

\acknowledgements

We are especially grateful to Emilio Falco for useful discussions 
and communications and John Tonry for helpful comments about his 
dynamical results. We also would like to thank Gary Bernstein for
providing the Hubble Space Telescope surface photometry of the
galaxy. The 4.2-m William Herschel Telescope is operated 
by the Isaac Newton Group, at the Spanish Observatorio del Roque de los 
Muchachos of the Instituto de Astrof\'\i sica de Canarias. We thank all 
the staff at the Observatory for their kind support.

This work was supported in part by DGESIC (Spain) grant 
PB97-0220-C02.


%
%

\clearpage

%
%

\thispagestyle{empty}
\begin{figure}

\caption{Slit superimposed on an image of 0957+561 obtained on the 
same night with the auxiliary Cassegrain camera. We have 
partially removed the A and B QSO images by using DAOPHOT.}

\end{figure}

\thispagestyle{empty}
\begin{figure}

\caption{Spectrum obtained co-adding all the exposures.}

\end{figure}

\thispagestyle{empty}
\begin{figure}

\caption{Continuum-corrected spectrum in the region of the Mg~{\sc i}{\it b} 
triplet. The spectrum has been blueshifted to $z=0$ (see text).}

\end{figure}

\thispagestyle{empty}
\begin{figure}

\caption{Calibration of the FWHM($\sigma_l$) relationship for 
three templates: the continuous, dotted, and dashed lines correspond to 
the AGK 2+14783, HD27697, and HD27697 smoothed star spectra, 
respectively.}

\end{figure}

\thispagestyle{empty}
\begin{figure}

\caption{Comparison of G1 spectrum with the spectrum 
corresponding to the calibration star AGK 2+14783 smoothed to 
$\sigma=310\rm km\,s^{-1}$}.

\end{figure}

\thispagestyle{empty}
\begin{figure}

\caption{$\chi^{2}_{\rm min}/dof$ (cross) and $\Delta(\chi^{2}/dof)$ 
= 1--10 contours (solid lines). We use the five dynamical data of 
Falco et al. (1997). The mass is expressed in terms of M$_{8}$ = 
10$^{8}$ $M_{\odot}$.}

\end{figure}

\thispagestyle{empty}
\begin{figure}

\caption{$\chi^{2}_{\rm min}/dof$ (cross) and $\Delta(\chi^{2}/dof)$ 
= 1--10 contours (solid lines). Our central 
stellar velocity dispersion (six dynamical constraints) has been included.
 The mass 
is expressed in terms of $M_{8}$ = 10$^{8}$ $M_{\odot}$.}

\end{figure}

\thispagestyle{empty}
\begin{figure}

\caption{$\chi^{2}_{\rm min}/dof$ (cross) and $\Delta(\chi^{2}/dof)$ 
= 1--10 contours (solid lines). We use all the dynamical data (Falco 
et al. + Tonry \& Franx + this paper). The mass is expressed in 
terms of $M_{8}$ = 10$^{8}$ $M_{\odot}$.}

\end{figure}

\thispagestyle{empty}
\begin{figure}

\caption{The ten measurements (with 1$\sigma$ error bars) of the 
stellar velocity dispersion (FSMD = filled circles, TF = open 
circles and this paper = open square) and the predicted values in 
two scenarios: $\sigma_h$ = 283 km s$^{-1}$, $hM_c$ = 10$^{9}$ 
$M_{\odot}$ (solid line) and $\sigma_h$ = 39 km s$^{-1}$, 
$hM_c$ = 1.9 x 10$^{10}$ $M_{\odot}$ (dashed line). See 
main text for details.}

\end{figure}

\thispagestyle{empty}
\begin{figure}

\caption{Surface brightness of the G1 isophotes is plotted vs. the 
mean radius of the isophotes. Filled triangles are the observational 
values in approximately the V band. Lines are different power laws
$I(\varphi) \propto \varphi^{-n}$: $n$ = 0.65 (dashed line from 0\farcs1
to 0\farcs2), $n$ = 1.26 (dashed line from 0\farcs2 to 5\arcsec) and
$n$ = 1.12 (solid line covering the whole range of radii). We note that
these indexes have been derived from {\it HST} observations by Bernstein
et al. (1997), and the slopes are different to the slope ($n \approx$ 2)
inferred from {\it KPNO} observations at larger radii (Bernstein, Tyson
\& Kochanek 1993).}

\end{figure}


\clearpage
\begin{figure*}[ht]
\epsfxsize=18 truecm
\centerline{\epsffile{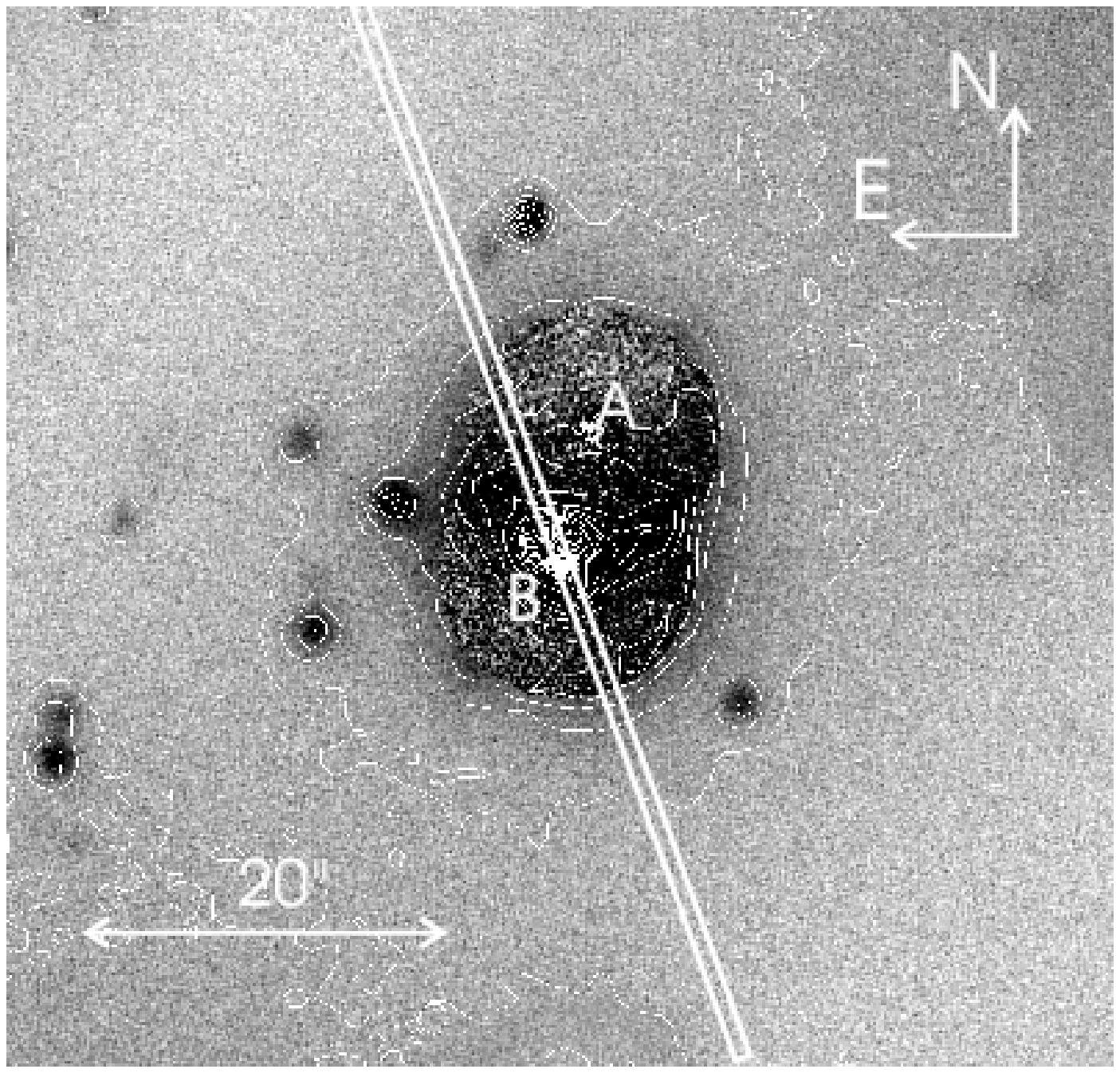}}
\end{figure*}
\clearpage
\begin{figure*}[ht]
\epsfxsize=18 truecm
\centerline{\epsffile{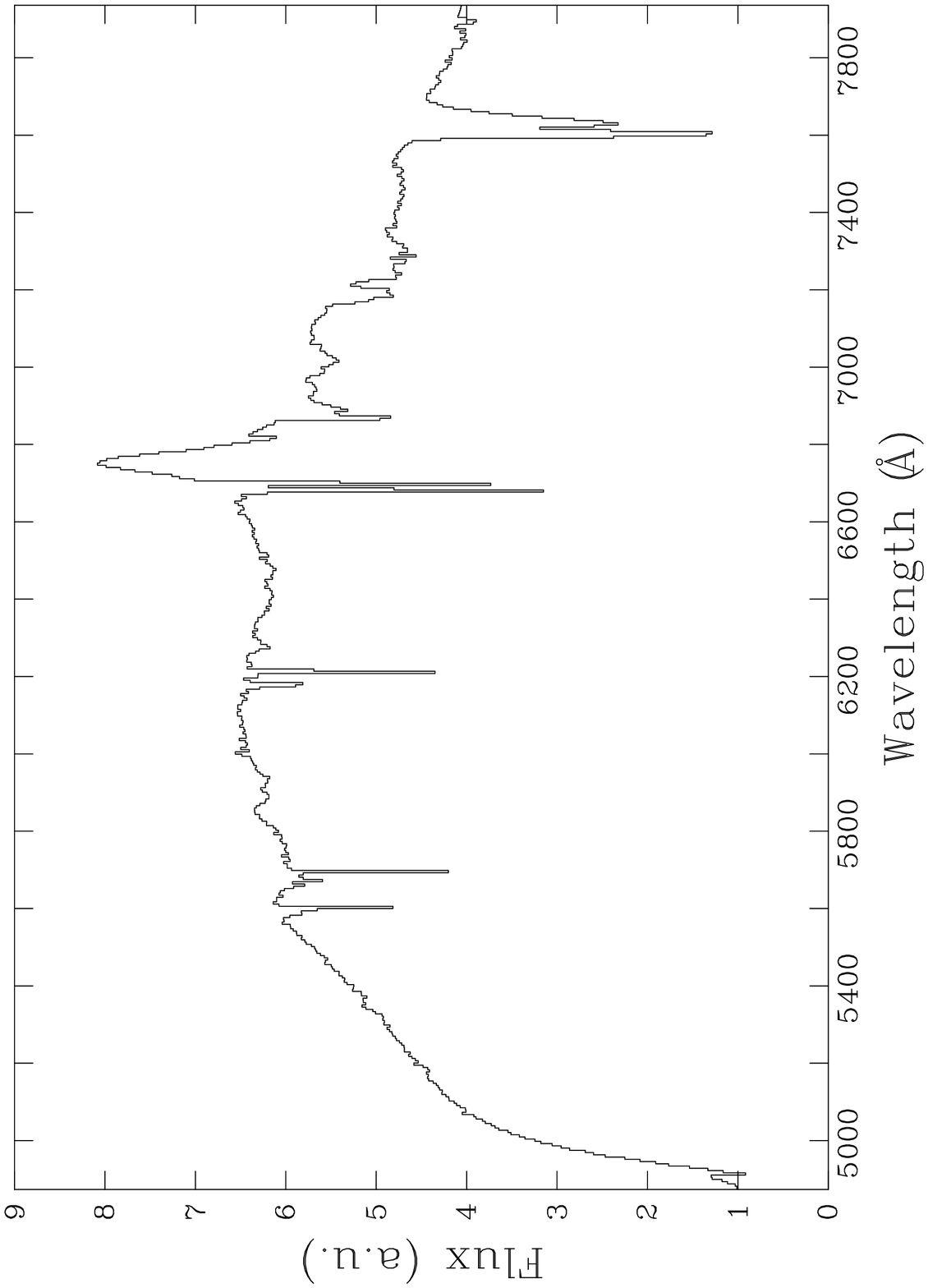}}
\end{figure*}
\clearpage
\begin{figure*}[ht]
\epsfxsize=18 truecm
\centerline{\epsffile{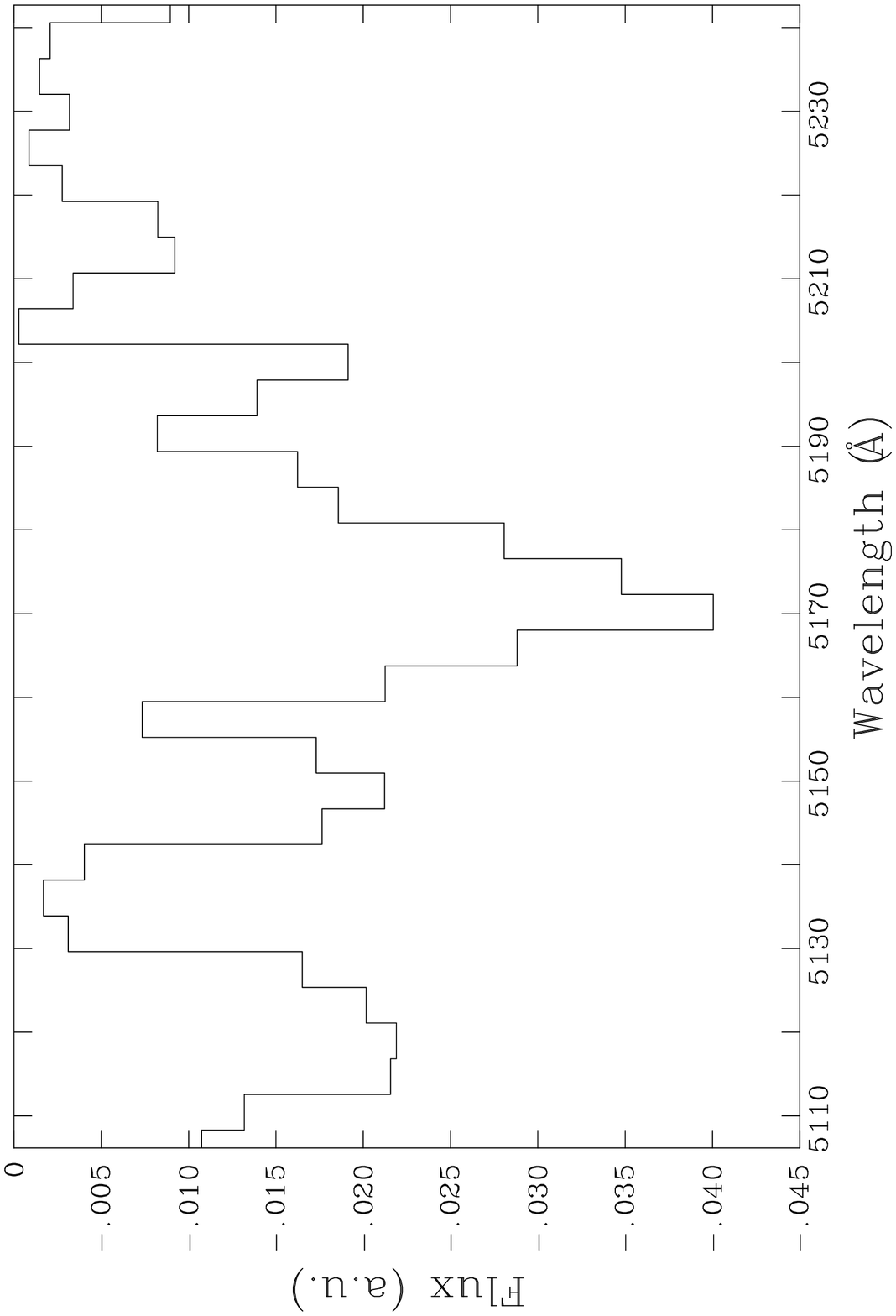}}
\end{figure*}
\clearpage
\begin{figure*}[ht]
\epsfxsize=18 truecm
\centerline{\epsffile{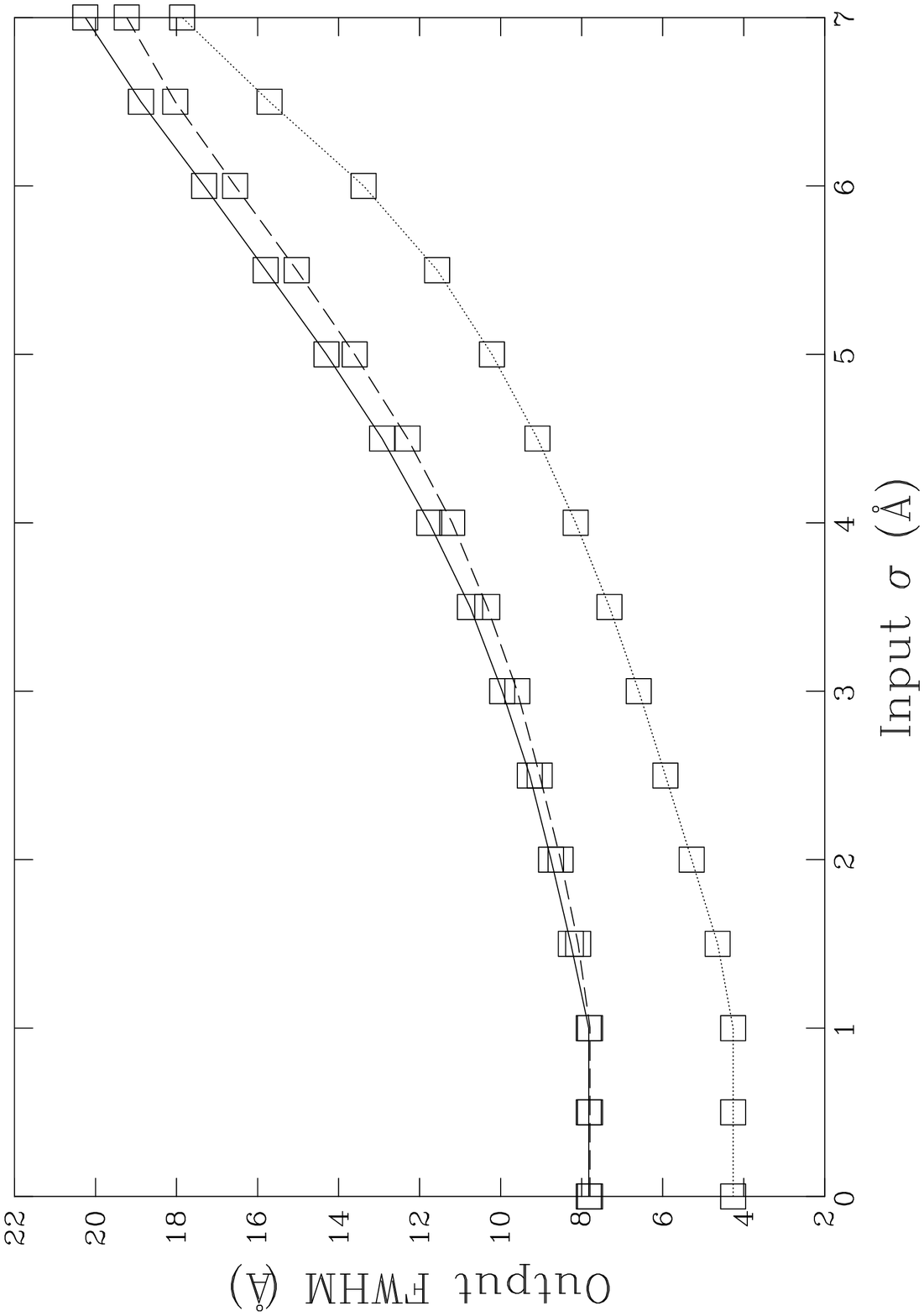}}
\end{figure*}
\clearpage
\begin{figure*}[ht]
\epsfxsize=18 truecm
\centerline{\epsffile{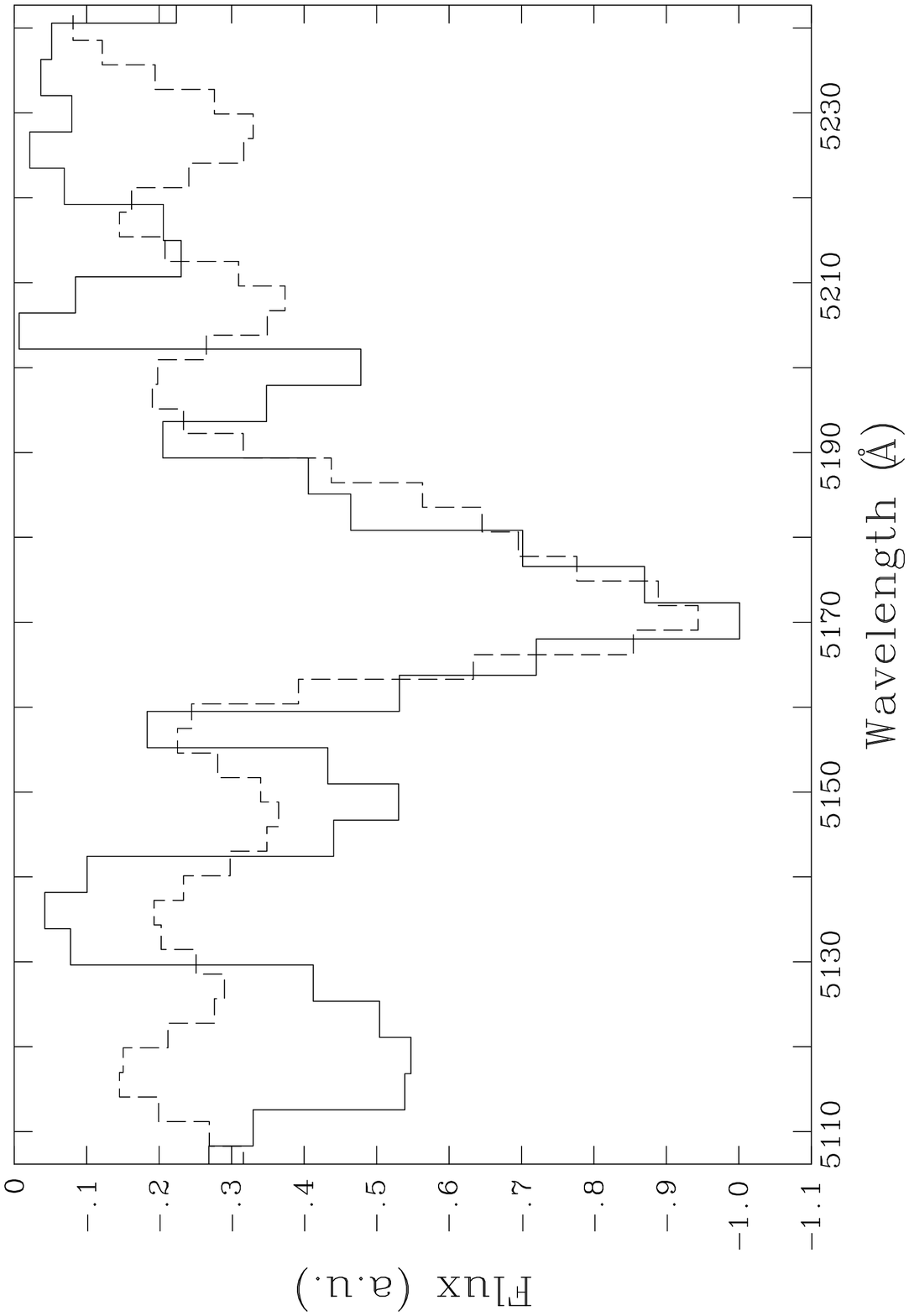}}
\end{figure*}
\clearpage
\begin{figure*}[ht]
\epsfxsize=18 truecm
\centerline{\epsffile{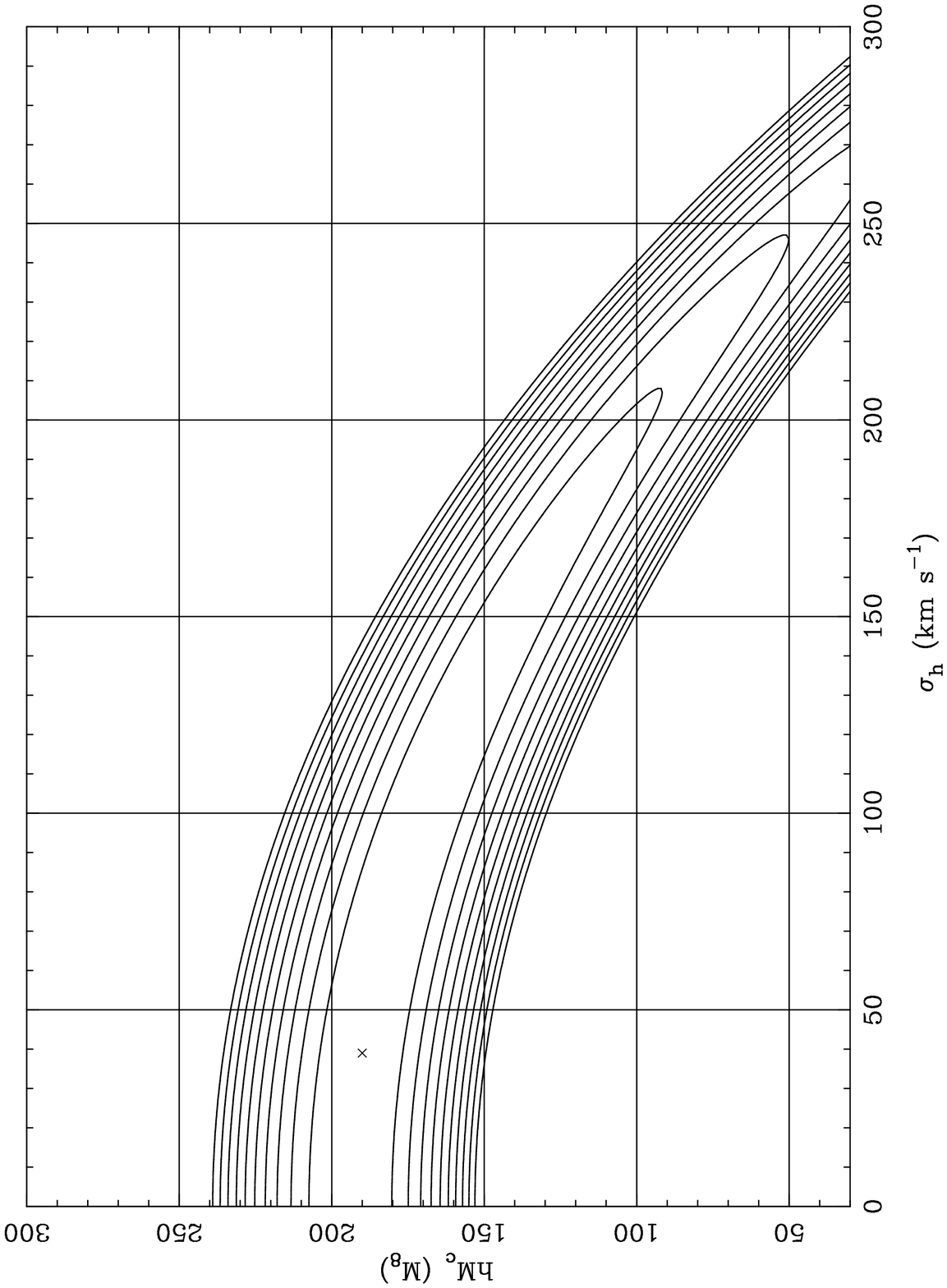}}
\end{figure*}
\clearpage
\begin{figure*}[ht]
\epsfxsize=18 truecm
\centerline{\epsffile{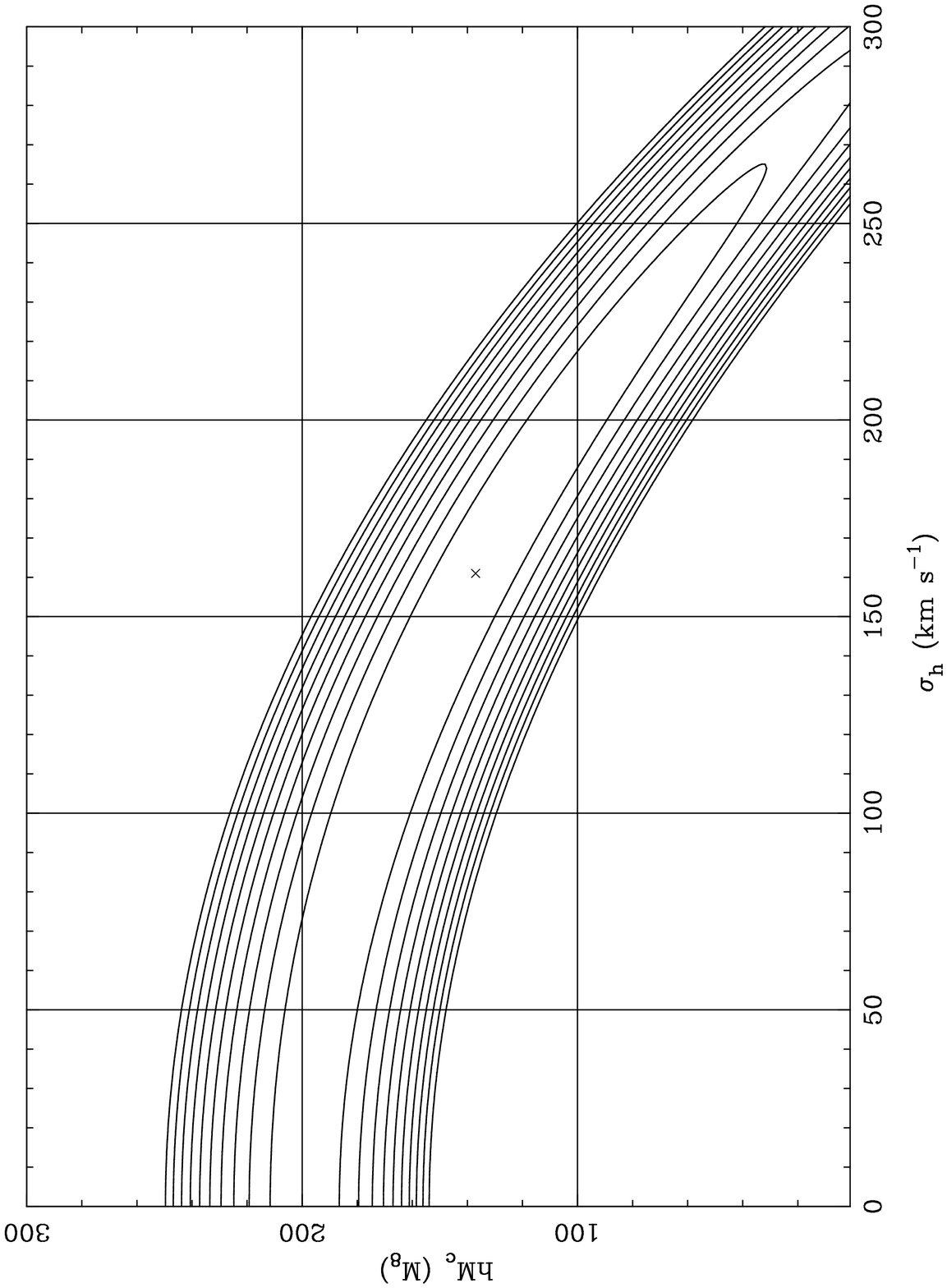}}
\end{figure*}
\clearpage
\begin{figure*}[ht]
\epsfxsize=18 truecm
\centerline{\epsffile{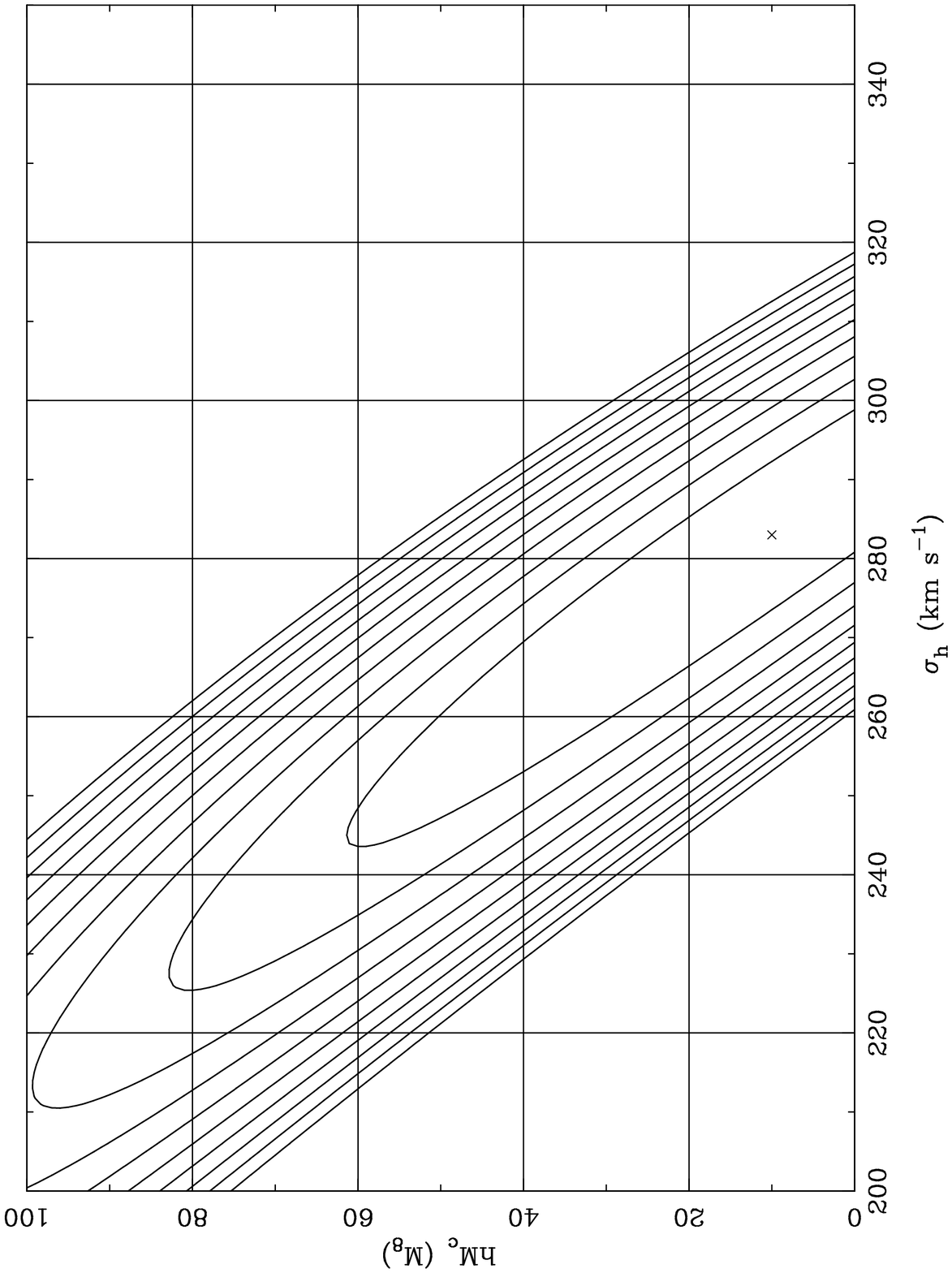}}
\end{figure*}
\clearpage
\begin{figure*}[ht]
\epsfxsize=18 truecm
\centerline{\epsffile{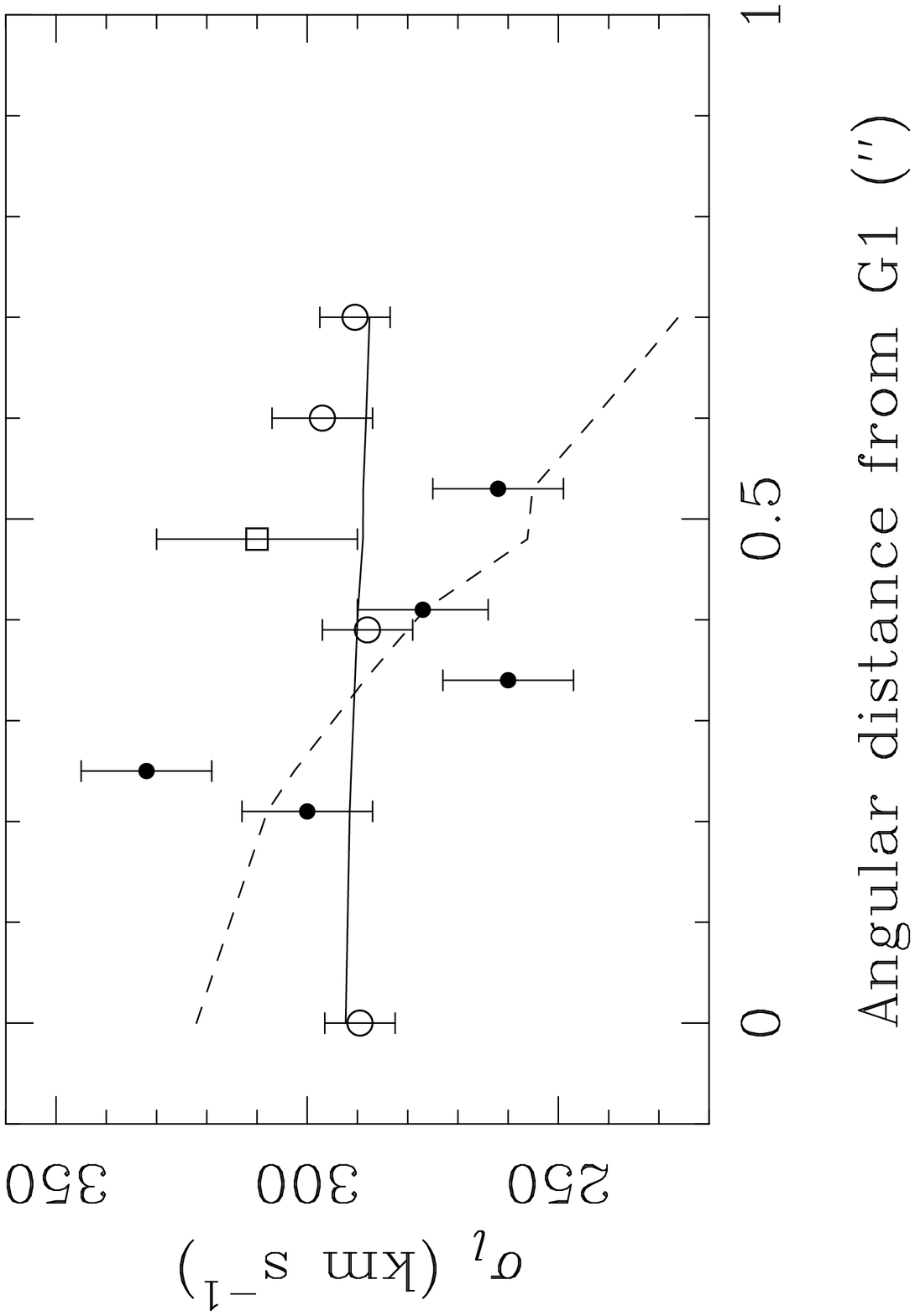}}
\end{figure*}
\clearpage
\begin{figure*}[ht]
\epsfxsize=18 truecm
\centerline{\epsffile{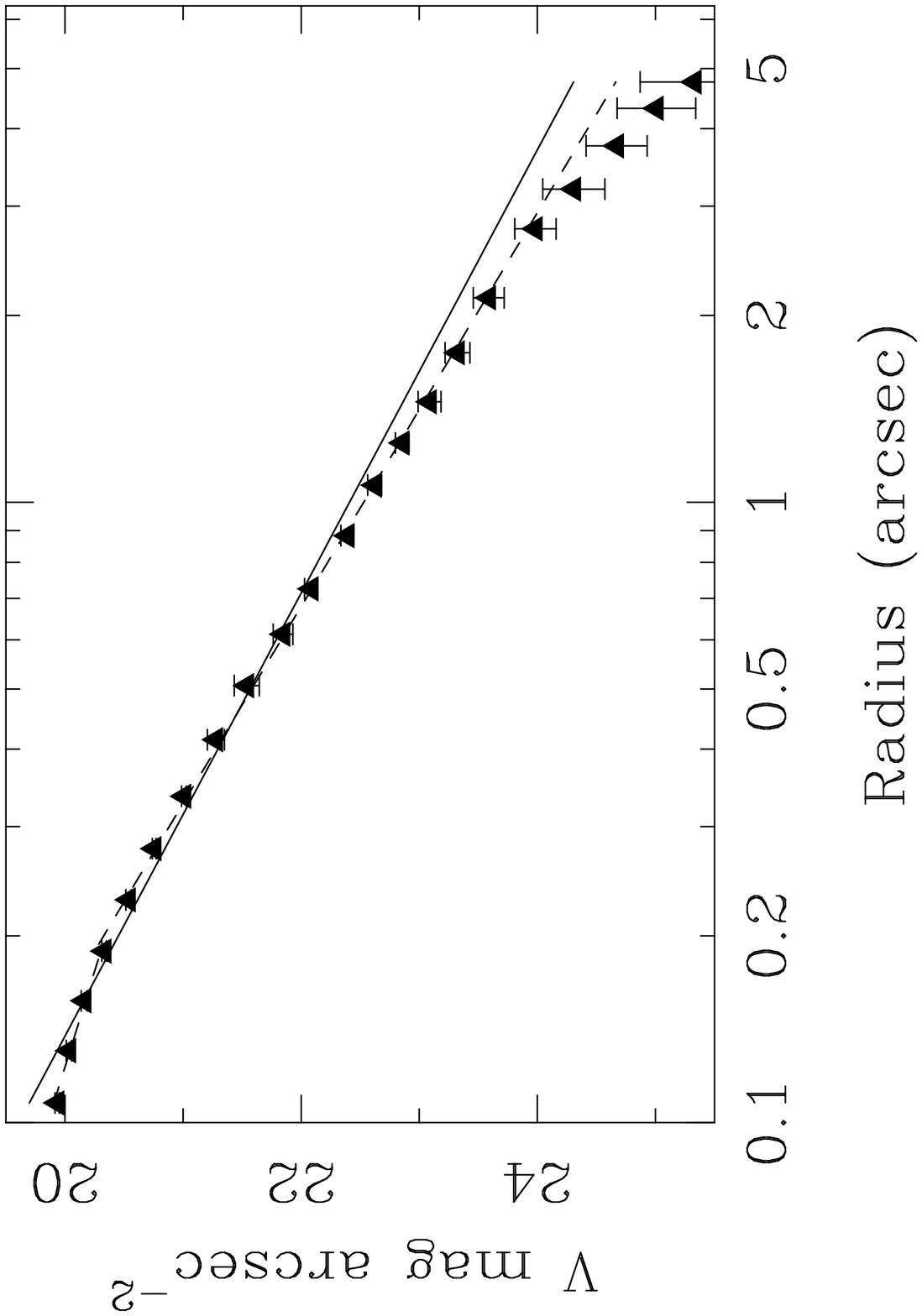}}
\end{figure*}

\end{document}